\documentclass{article}
\usepackage{spconf,amsmath,graphicx}
\usepackage{url}
\usepackage{multirow}

\title{Improving Short Utterance Anti-Spoofing with AASIST2}
\name{Yuxiang Zhang$^{12}$, Jingze Lu$^{12}$, Zengqiang Shang$^{1}$, Wenchao Wang$^{1}$, Pengyuan Zhang$^{12}$\thanks{Corresponding author: Wenchao Wang and Pengyuan Zhang. This work is supported in part by Youth Innovation Promotion Association CAS and by National Nature Science Foundation of China (12204509).}}
\address{$^1$Key Laboratory of Speech Acoustics and Content Understanding, Institute of Acoustics, \\Chinese Academy of Sciences, Beijing, China\\$^2$University of Chinese Academy of Sciences, Beijing, China}
\begin{document}
%
\maketitle
\begin{abstract}
The wav2vec 2.0 and integrated spectro-temporal graph attention network (AASIST) based countermeasure achieves great performance in speech anti-spoofing. However, current spoof speech detection systems have fixed training and evaluation durations, while the performance degrades significantly during short utterance evaluation. To solve this problem, \mbox{AASIST} can be improved to \mbox{AASIST2} by modifying the residual blocks to Res2Net blocks. The modified Res2Net blocks can extract multi-scale features and improve the detection performance for speech of different durations, thus improving the short utterance evaluation performance. On the other hand, adaptive large margin fine-tuning (ALMFT) has achieved performance improvement in short utterance speaker verification. Therefore, we apply Dynamic Chunk Size (DCS) and ALMFT training strategies in speech anti-spoofing to further improve the performance of short utterance evaluation. Experiments demonstrate that the proposed \mbox{AASIST2} improves the performance of short utterance evaluation while maintaining the performance of regular evaluation on different datasets.
\end{abstract}
\section{Introduction}
With the recent surge of Artificial Intelligence Generated Content (AIGC), spoofing algorithms have also gained momentum. The convenience and the quality of generating spoof speech  have improved significantly. Consequently, there is an increased risk of malicious use of spoof speech. Spoof speech is now used not only to attack automatic speaker verification (ASV) systems but also for telecommunications fraud and cognitive warfare. Much work has been proposed to prevent the dangers of spoof speech, facilitated in particular by the flagship ASVspoof Challenge series~\cite{wu2015asvspoof, Kinnunen2017, todisco2019asvspoof, yamagishi21_asvspoof}. The latest ASVspoof 2021 Challenge considered the impact of cross-channel and compression codecs on spoof speech detection systems. The spoof speech countermeasure (CM) based on pre-trained wav2vec 2.0 \cite{baevski2020wav2vec} and an integrated spectro-temporal graph attention network (AASIST)~\cite{tak22_odyssey} achieves good results on multiple datasets~\cite{wang2023spoofed}.

However, a common problem in speech signal processing: the degradation of short-term speech evaluation performance has not been fully explored in spoof speech detection. The training and evaluation of current spoof speech CMs typically require the speech segments of four seconds or longer~\cite{lavrentyeva2019stc, 9747766, zhang23ca_interspeech}. And since the duration of a batch of input speech is fixed during training, the speech is truncated or repeated. Mismatches in length between training and evaluation, as well as between different datasets also lead to performance degradation~\cite{rosello23_interspeech}. Similar to the ASV systems, increasing the audio length during evaluation can improve the performance of the spoof speech detection systems~\cite{wang2023low}. However, as human-computer interaction via speech becomes more common, there is an urgent need to address the accurate detection of spoof speech in short segments in the field of short utterance ASV such as smart homes, smart cars, voice wake-up and online banking authentication.

There has been some research in the area of ASV to improve the performance of short speech evaluation \cite{hajavi19_interspeech, kanagasundaram19_interspeech}. Res2Net~\cite{gao2019res2net} has shown good performance in both ASV~\cite{zhou2021resnext} and speech anti-spoofing \cite{li2021replay} field. The concept of dividing features into distinct groups within a block and extracting multi-scale features is also utilized in popular models such as ECAPA-TDNN \cite{desplanques20_interspeech}. Since extracting multi-scale features in Res2Net blocks expands the receptive field of the model, it has the potential to obtain richer information from short utterances, thus improving the performance of short-term evaluation. In addition, duration-based adaptive large margin fine-tuning (ALMFT) \cite{zhang2023adaptive} is proposed to improve the robustness of short-duration ASV. The training strategy considers the varying classification difficulties associated to different speech lengths during the training process. However, it has not been applied to speech anti-spoofing. Therefore, the dynamic chunk size (DCS) and the ALMFT training strategy is introduced in the training of \mbox{AASIST2}. 

Based on the previous work, five ResNet blocks in AASIST are modified to the Res2Net blocks, and the Additive Angular Margin Softmax (AM-Softmax) loss function~\cite{deng2019arcface} is introduced, resulting in AASIST2. To further improve the robustness of CM for different duration evaluations, the DCS and ALMFT training strategies are introduced. Experimental results demonstrate that the proposed CM exhibits state-of-the-art (SOTA) performance over different evaluation datasets for different speech lengths.

\section{Method}
This section describes the proposed AASIST2 model and the different training strategies used to improve the performance of short utterance evaluation in detail.
\subsection{AASIST2}
AASIST consists of two parts: a RawNet2-based encoder~\cite{jung20c_interspeech} and a graph network-based module. In the original RawNet2 and AASIST models, the encoder extracts 2-dimensional (2D) features directly from the speech waveform via a sinc-convolution layer. Replacing the sinc layer with pre-trained wav2vec 2.0 for feature extraction effectively improves the performance and robustness of AASIST. The extracted features are fed into a 2D residual network containing six residual blocks to learn higher-level feature representations. Each residual block consists of two 2D convolutional layers, SeLU activation function, batch normalization (BN), and max-pooling. Although the current residual network already has a good anti-spoofing performance, Res2Net can achieve a significant improvement when dealing with speech features that are rich in temporal information.

\begin{figure}[htb]
	
	\begin{minipage}[b]{.32\linewidth}
		\centering
		\centerline{\includegraphics[width=2.6cm]{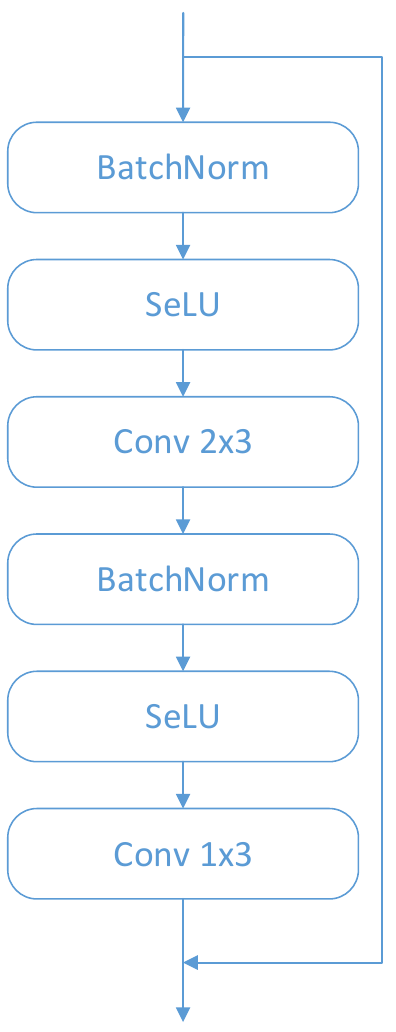}}
		\centerline{(a) Residual Block}\medskip
	\end{minipage}
	\hfill
	\begin{minipage}[b]{0.64\linewidth}
		\centering
		\centerline{\includegraphics[width=5.2cm]{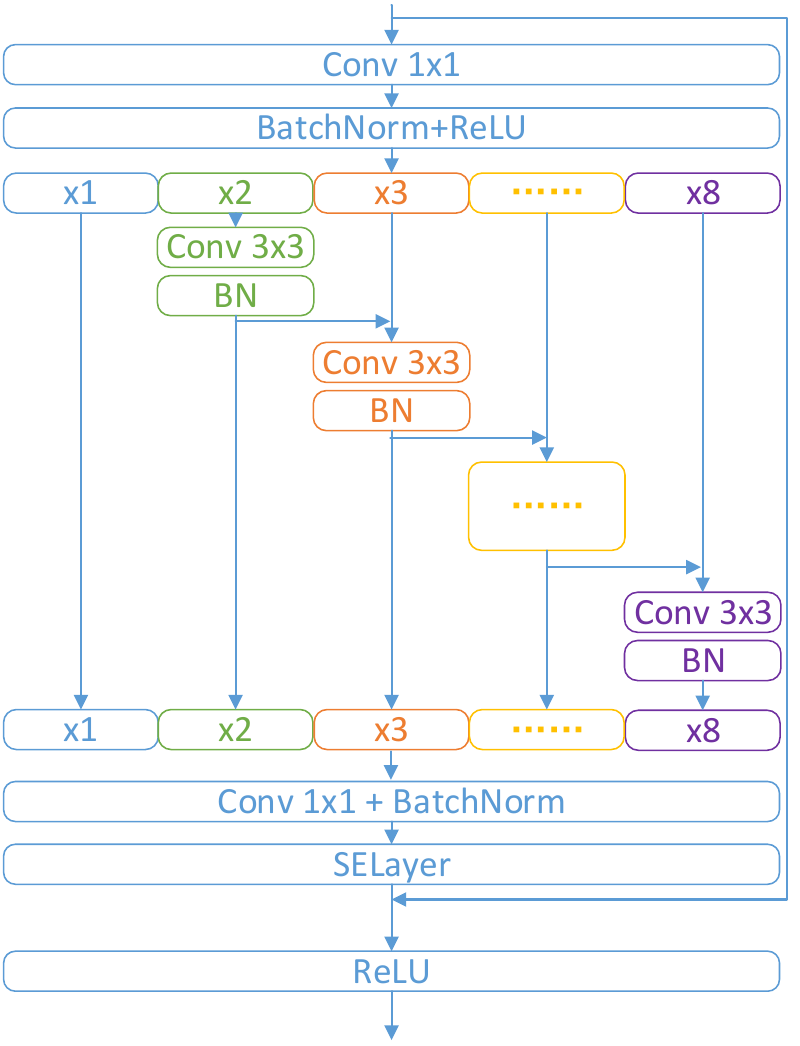}}
		\centerline{(b) Res2Net Block}\medskip
	\end{minipage}
	\caption{The illustration of the residual block in AASIST and Res2Net block in AASIST2.}
	\label{fig:res}
\end{figure}

The Res2Net block enlarges the receptive fields trough grouping features and hierarchical convolution to extract multi-scale representations. The residual blocks before and after the modification in AASIST are illustrated in Fig. \ref{fig:res}. The Res2Net block averages the input feature in channel dimensions into $s$ groups, denoted by $x_i$ , where $i \in {1, 2, ... , s}$. Each $x_i$ except $x_1$ is separately filtered by a $3 \times 3$ convolutional filter $K_i$. Starting from $i = 3$, $x_i$ is summed with the output of $K_{i-1}$ , which is then fed into $K_i$. The following Eq. \ref{eq1} shows the forward process of the Res2Net block:
\begin{equation}\label{eq1}
	\begin{aligned}
		y_i = \begin{cases}
			x_i, & i=1\\
			K_i(x_i), & i=2\\
			K_i(x_i + y_{i-1}), & 2<i\le s
		\end{cases}
	\end{aligned} 
\end{equation}
The hierarchical connections similar to residual networks between different groups of features can alleviate the gradient vanishing problem and facilitate training. Additively, the last squeeze-and-excitation (SE) layer of the Res2Net block can assign weights to the features of different channels through attention, thereby enhancing the features that are more important for speech anti-spoofing.

In addition, AM-Softmax is used as the loss function for the training of AASIST2. In the original AASIST, the loss function is weighted cross-entropy loss. The AM-Softmax loss function improves the ability of the model to discriminate between bonafide and spoof speech by increasing the interclass distance. Moreover, the large margin and feature normalization make the outliers have less impact on the model, which improves the model generalization and robustness. The AM-Softmax loss function is calculated as follows Eq. \ref{eq2} :
\begin{equation}\label{eq2}
	\mathcal{L}_{AMS} = -\frac{1}{n} \sum_{i=1}^{n}\log{\frac{e^{s\cdot\left(\boldsymbol{W}_{y_i}^T \boldsymbol{f}_i-m\right)}}{e^{s\cdot\left(\boldsymbol{W}_{y_i}^T \boldsymbol{f}_i-m\right)}+\sum_{j=1,j\ne y_i}^{c}e^{s \boldsymbol{W}_j^T \boldsymbol{f}_i}}}
\end{equation}
where $\boldsymbol{f}$ is the input of the last fully connected layer, $\boldsymbol{W}$ is the weight matrix of the last fully connected layer, $s$ is the scale factor and $m$ is the margin.

Expanding the receptive field by extracting multi-scale features through Res2Net blocks improves the capacity of modeling training speech of different durations. And SELayer can enhance the feature channels with stronger discriminative ability. Meanwhile, AM-Softmax loss further strengthens the discriminative ability of AASIST2 and its robustness to abnormal samples. These improvements enable AASIST2 to perform better not only for regular evaluations but also for short utterance evaluations.

\subsection{Dynamic Chunk Size and Large Margin Fine-Tuning}
The speech in the dataset varies in duration, but the size of each batch of input speech must be the same during training. Therefore, it is necessary to fix the input speech to a uniform length using methods such as truncation or padding. In speaker recognition and language recognition, variable-length training produces better results than fixed-length training \cite{9036861}. Therefore, a similar DCS training strategy is introduced into the training of the proposed AASIST2. In the training process, a random integer $N$ is first generated from the uniform distribution in the interval $(N_{min}, N_{max})$ as the chunk size for each batch. Then the length of each speech is truncated or padded to $N$ when the data is fed into this batch.

Furthermore, increasing the duration of the test speech while keeping the duration of the training speech fixed can improve the performance of the CM. So it can be assumed that longer speech is less difficult to discriminate. And increasing the margin in AM-Softmax can make classification more difficult. Therefore, it is possible to increase the margin for long speech to improve the model discrimination ability when training the model, which is called large margin fine-tuning (LMFT). LMFT is a secondary training phase based on network models that have already converged to help create more robust embeddings. In order to keep the system stable at difficult large margin settings, longer training utterances are fed to provide more information. As wav2vec 2.0 is a pre-trained model, the subsequent training is a fine-tuning phase. Instead of conducting a secondary phase of finetune, here we change the margin during the model training with DCS.

Spoof speech of different durations is discriminated with different difficulties, thus in order to combine LMFT with DCS, adaptive margins can be applied for different speech durations, resulting in ALMFT. A dynamic function that allows the margins to adapt to the duration of the training speech is introduced as a linear transformation. The formula characterizing the increasing relationship between the margin and the speech duration is as follows:

\begin{equation}\label{eq3}
	\mathrm{Margin} = A \times \mathrm{Duration} + B
\end{equation}
where the values of $A$ and $B$ in the linear transformation can be parameterized by setting a range of margin values and a range of speech durations. And the range of speech durations is $\mathrm{Duration} \in (N_{min}/sr, N_{max}/sr)$, while $sr$ is the sample rate of speech.

\section{Experimental Setup}
The datasets, metrics, model details and training parameters utilized for the experiments are described in this section.

\subsection{Datasets and Metrics}

The training of the AASIST2 is based on the training set of the ASVspoof 2019 LA database. And the development partition is used to select the models. The model with the lowest loss on the development set is selected for evaluation. The evaluation data includes the evaluation partition of the ASVspoof 2019 LA (19), ASVspoof 2021 LA (LA), as well as ASVspoof 2021 DF (DF) datasets. 

The metric used here is the Equal Error Rate (EER), which is the evaluation metric used by ASVspoof Challenges.

\subsection{Model Details}
The model based on wav2vec 2.0 and AASIST2\footnote{\url{https://github.com/TakHemlata/SSL_Anti-spoofing}} follows the example in \cite{tak22_odyssey} except for the encoder and loss function. The pre-trained wav2vec 2.0 XLS-R is utilized for feature extraction. The features are reduced in dimension from 1024 to 128 by a linear layer and fed into the subsequent \mbox{AASIST2}. 

In the encoder of AASIST2, except for the first residual block, the remaining five residual blocks are modified as Res2Net blocks. The Res2Net blocks have a width of 14 and a scale of 8. The scale of AM-Softmax is 15, and the margin changes with the dynamic chunk size adaptively. 

In the example of \cite{tak22_odyssey}, different RawBoost\footnote{\url{https://github.com/TakHemlata/RawBoost-antispoofing}} methods are adopted for the ASVspoof 2021 LA and DF datasets to achieve better results. Benefiting from the improvements of AASIST2, the configuration $algo=4$ with all three \mbox{RawBoost} filters is applied here for the data augmentation for all evaluation datasets.

\subsection{Training Strategy}
For regular training, speech is truncated or concatenated to about 4s (64,600 samples). For DCS, the length of speech is chosen randomly within the interval from 1s to 6s, and the number of speech samples ranges from 16,000 to 96,000. For ALMFT, the value of margin is in the range of $0.2$ to $0.5$. So the parameters in Eq. \ref{eq3} are $A =\frac{3}{50}$ and $B=\frac{7}{50}$.

The Adam optimizer with a fixed learning rate of $10^{-6}$ and batch size of 16 are utilized for experiments. Parameters of the front-end wav2vec 2.0 feature extractor and the back-end AASIST2 model are jointly optimized. All models are trained for 100 epochs on a single GeForce A10 GPU.

\section{Results and Analysis}
This section analyzes the experimental results and demonstrates the effectiveness of AASIST2, DCS, and LAMFT.

\begin{table*}[htbp]
	\caption{EER/\% under evaluation of different speech durations. DCS: proposed dynamic chunk size; ALMFT: proposed adaptive large margin fine-tuning with DCS. Optimal results for different datasets at different evaluation durations are \textbf{bolded}.} \label{tab1}
	\centering
	\setlength{\tabcolsep}{2.3pt}
	\begin{tabular}{ l | c c c | c c c | c c c | c c c | c c c | c c c | c c c}
		\hline
		\multirow{2}{*}{Model} & \multicolumn{3}{c}{1s} & \multicolumn{3}{c}{2s} & \multicolumn{3}{c}{3s} & \multicolumn{3}{c}{4s} & \multicolumn{3}{c}{5s} & \multicolumn{3}{c}{6s} & \multicolumn{3}{c}{Variable-length} \\
		\cline{2-22}
		 & 19 & LA & DF & 19 & LA & DF & 19 & LA & DF & 19 & LA & DF & 19 & LA & DF & 19 & LA & DF & 19 & LA & DF\\
		\hline
		Baseline & 10.1 & 12.5 & 12.1 & 1.88 & 3.56 & 5.66 & 0.73 & 1.94 & 4.37 & 0.21 & 1.76 & 4.38 & 0.18 & 1.58 & 4.57 & 0.18 & 1.42 & 4.84 & 0.16 & 2.18 & 4.67 \\
		+ DCS & 9.14 & \textbf{10.4} & 10.2 & 1.81 & \textbf{2.49} & 6.54 & 0.75 & \textbf{1.48} & 6.76 & 0.39 & \textbf{1.08} & 7.14 & 0.26 & \textbf{0.90} & 7.37 & 0.20 & \textbf{0.86} & 7.80 & 0.18 & \textbf{1.13} & 8.02 \\
		\hline
		AASIST2 & 9.69 & 11.7 & \textbf{5.78} & 1.66 & 3.37 & \textbf{3.50} & \textbf{0.57} & 2.13 & \textbf{2.88} & 0.22 & 1.76 & \textbf{2.62} & 0.16 & 1.64 & \textbf{2.60} & 0.15 & 1.61 & \textbf{2.77} & 0.83 & 2.50 & \textbf{2.04} \\
		+ ALMFT & \textbf{8.36} & 10.7 & 8.55 & \textbf{1.44} & 3.21 & 5.21 & 0.63 & 1.81 & 4.39 & \textbf{0.22} & 1.39 & 4.06 & \textbf{0.15} & 1.30 & 3.89 & \textbf{0.12} & 1.33 & 3.75 & \textbf{0.16} & 1.59 & 4.59 \\
		\hline
	\end{tabular}
	
\end{table*}

\subsection{Effectiveness of AASIST2}
The baseline system is the same as \cite{tak22_odyssey} except $algo=4$. Comparing the results of the baseline system and AASIST2 both trained and evaluated at speech length of 4s in Table \ref{tab1}, it can be found that AASIST2 achieves significant performance improvements over the baseline system on all three datasets. In particular, on the ASVspoof 2021 DF dataset, the EER is relatively reduced by 40.2\%. And AASIST2 trained by ALMFT achieves optimal results in tests of different durations on the ASVspoof 2019 LA dataset. A possible reason for this is that the Res2Net blocks extract multi-scale features and thus achieves better robustness.

Table \ref{tab2} compares the results of the AASIST2 with the current SOTA systems. AASIST2 achieves SOTA results on the ASVspoof 2019 LA evaluation partition. On the ASVspoof 2021 LA and DF evaluation sets, AASIST2 does not select the appropriate data augmentation method based on the dataset as in \cite{tak22_odyssey, rosello23_interspeech} and still achieves comparable results.

\begin{table}[htbp]
	\caption{Performance comparison with SOTA systems.} \label{tab2}
	\centering
	\begin{tabular}{ l c c c }
		\hline
		System & 19 & LA & DF \\
		\hline
		DFSincNet \cite{huang2023discriminative} & 0.52 & 3.38 & - \\
		Wav2vec + FC \cite{wang2023spoofed} & 0.21 & 3.30 & 4.12 \\
		Wav2vec + AASIST \cite{tak22_odyssey} & - & 0.82 & 2.85 \\
		Wav2vec + Conformer \cite{rosello23_interspeech} & - & 0.97 & 2.58 \\
		\hline
		Wav2vec + AASIST2 (Ours) & \textbf{0.15} & 1.61 & 2.77 \\
		\hline
	\end{tabular}
\end{table}

\subsection{Improvement of Short Utterance Evaluation}
\begin{figure}[htb]
	\centering
	\includegraphics[width=1.0\linewidth]{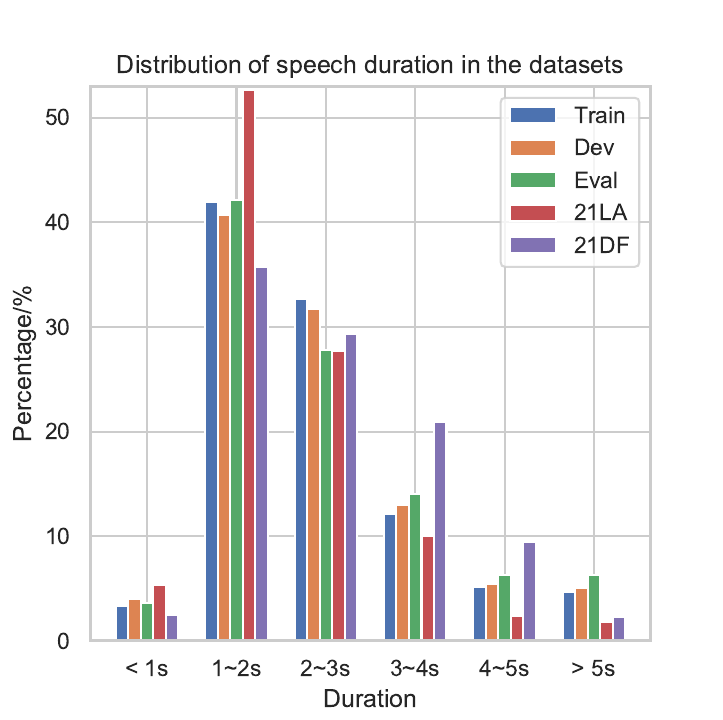}
	\caption{The distribution of speech duration in the datasets.}
	\label{fig:bar}
\end{figure}

With the experimental results of different speech duration evaluations in Table \ref{tab1}, AASIST2, DCS, and ALMFT all achieve performance gains in the short utterance evaluation. The EERs of all systems are substantially larger in the evaluation of 1s. AASIST2 with fixed duration training has the worst performance on the ASVspoof 2019 and ASVspoof 2021 LA datasets but achieves optimal results after ALMFT. Training with ALMFT results in a relative reduction in EER of 13.7\% and 8.5\% on the ASVspoof 2019 LA and ASVspoof 2021 LA datasets compared to fixed duration training. However, there is a significant performance degradation on the ASVspoof 2021 DF dataset. In order to analyze the reasons, we statistically analyze the distribution of speech durations in different datasets, and the results are shown in Fig \ref{fig:bar}.

As shown in Fig \ref{fig:bar}, the ASVspoof 2021 DF dataset has significantly less short duration speech ($<2$s) compared to the other datasets, and more speech in length of 3s to 5s. This may result in systems trained with speech of 4s performing better on the ASVspoof 2021 DF dataset, while dynamically varying training durations and margins result in degraded performance on short utterance tests. And the performance degradation due to ALMFT in long utterance evaluation is relatively small.

While there is no margin effect for the baseline system and DCS, the dynamic training duration increases the robustness to short utterance tests of 1s. And the optimal performance is achieved on the ASVspoof 2021 LA dataset with more short utterances ($<2$s). However, as the length of the test speech increases, DCS loses its advantage except for the ASVspoof 2021 LA dataset. This is probably because the ASVspoof 2021 DF has more long utterances. There is little difference in performance on the ASVspoof 2019 LA dataset. 

\section{Conclusion}
In this paper, the AASIST2 model is obtained by modifying the residual blocks in AASIST into Res2Net blocks and introducing AM-Softmax loss. The performance of anti-spoofing is further improved based on the multi-scale feature extraction and large margin. Furthermore, the effect of speech duration mismatch on training strategies is analyzed. And with the dynamic chunk size and ALMFT training strategies, the performance of both the baseline system and the AASIST2 system for short utterance evaluation is effectively improved.

\newpage
\bibliographystyle{IEEEbib}
\bibliography{mybib}

\end{document}